\documentclass{article}

\usepackage[english]{babel}
\usepackage[a4paper,top=2.5cm,bottom=3cm,left=3.5cm,right=3.5cm,marginparwidth=1.75cm]{geometry}
\usepackage[T1]{fontenc}
\usepackage[latin1]{inputenc}
\usepackage{graphicx}
\usepackage{amsfonts, amsmath}
\usepackage{amssymb,mathtools}
\usepackage{amstext}
\usepackage{color}
\usepackage{dsfont}
\usepackage{subfig}
\usepackage{parskip}

\begin{document}
\graphicspath{{figs/}}

\title{Adaptation and synchronization -- \\basic mechanisms in music performance}

\author{Jakub Sawicki$^{1,2,3}$}
\date{
    $^1$Potsdam Institute for Climate Impact Research\\%
    $^2$University of Applied Sciences Northwestern Switzerland\\
    $^3$Berlin University of the Arts\\[2ex]%
    \today
}

\maketitle

\begin{abstract} 
This review examines the roles of adaptation and synchronization in music performance, drawing on concepts from complex systems theory to understand the dynamic interactions between musicians, music, and listeners. Adaptation is explored through how musicians adjust their cognitive, emotional, and motor systems across the stages of preparation, execution, and reception, while synchronization is emphasized as essential for aligning internal states, coordinating actions with other performers, and engaging with the audience. The review highlights the interdisciplinary nature of music performance research, integrating cognitive, motor, and emotional processes to enhance both individual and collective musical expression. It also addresses the psychological state of flow, which arises from synchronized neurocognitive mechanisms that optimize performance. Additionally, the emotional synchronization facilitated by music is explored, emphasizing its role in both individual emotional coherence and social coordination within musical ensembles. Finally, the review highlights recent findings on interpersonal and inter-brain synchronization, particularly in live music performances and improvisation, showing how synchronization fosters creativity, social cohesion, and a shared collective experience.
\end{abstract}

\section{The interdisciplinary nature of music performance}

In this review, we examine the multifaceted interplay among the musician, the music, and the listener within the context of live and recorded music performance. This tripartite relationship encompasses the physiological, psychological, and neuronal development of the musician, the structural and expressive features of the music, and the perceptual and affective responses of the audience. Central to this interaction are the mechanisms of adaptation and synchronization, which mediate coordination within and across individuals, enabling shared experiences and effective performance outcomes. Although the examples discussed herein are not exhaustive, they serve to illustrate how these mechanisms underpin key phases of the performance process.

\begin{figure}
\centering
\includegraphics[width=.65\linewidth]{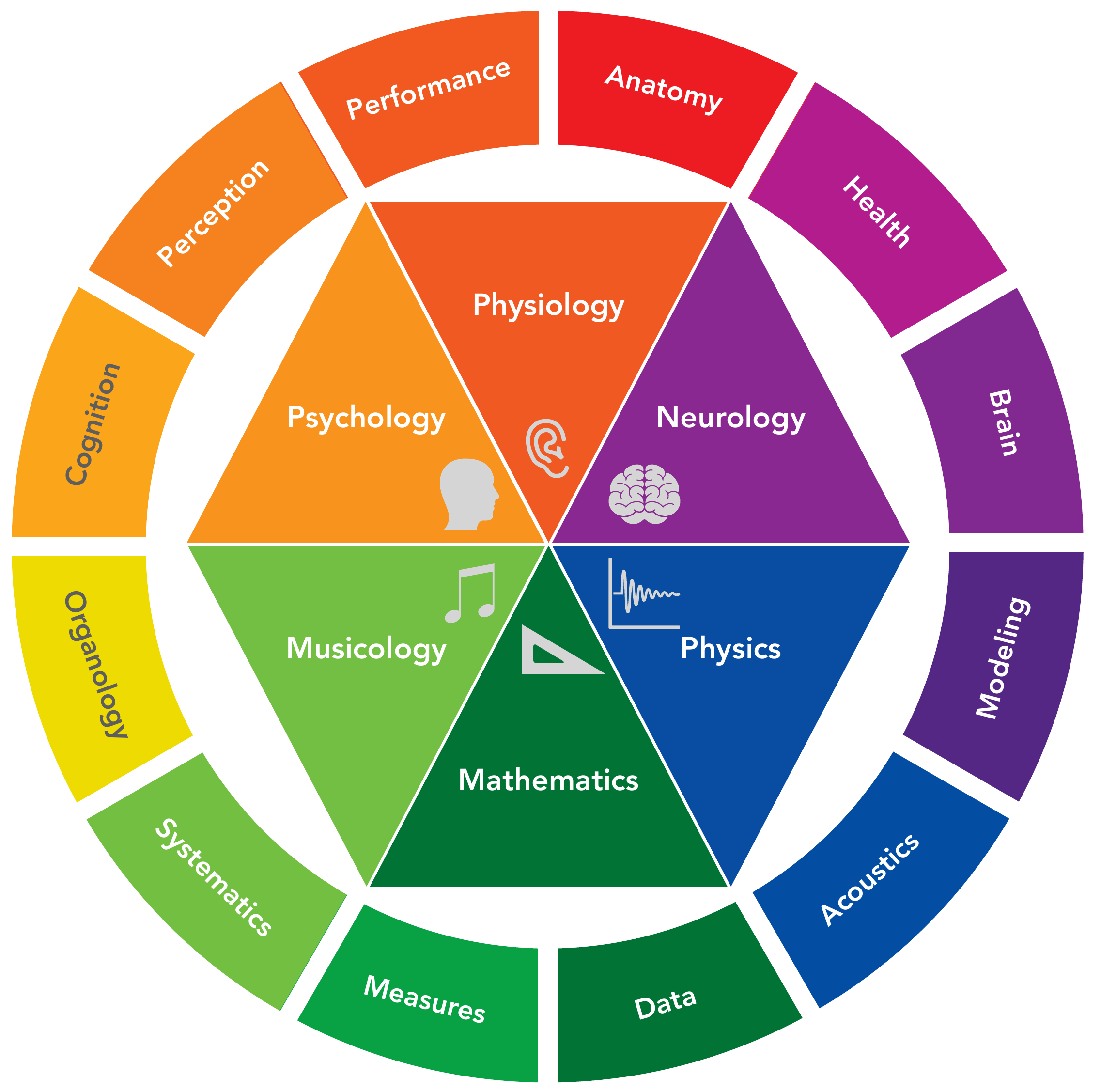}
\caption{The figure illustrates the balance of diverse inter- and transdisciplinary connections within the field of music psychology, highlighting the integrative position of music performance studies. Central to the framework are the contributing scientific disciplines, which provide foundational theories and methodologies. Surrounding these core domains are their respective applied contexts, reflecting the translation of theoretical insights into practical applications across research and performance settings.
\label{intro:fig:musicpsychology}
}
\end{figure}

As depicted in Fig.\,\ref{intro:fig:musicpsychology}, music performance studies has emerged as an interdisciplinary subfield within music psychology \cite{PAL97}, integrating methodologies from the cognitive sciences, neuroscience, ethnomusicology, and performance studies. The field conceptualizes performance both as a dynamic process and as an artifact of artistic creation. Methodologically, it embraces both practice-led and practice-based research approaches, combining empirical investigation with performance as a mode of inquiry \cite{MCP22, MCP22a}. The inherently interdisciplinary character of this field complicates its strict definition but enables a comprehensive examination of how musical ideas are formed, embodied, and communicated.

Musical performance is not merely a medium of aesthetic expression but a fundamental component of human culture and communication. Its origins are intimately tied to physiological rhythms -- such as heartbeat, breathing, and vocalizations associated with emotional expression -- suggesting an evolutionary basis for musical behavior. These primitive expressions evolved into more structured musical forms, supported by archaeological findings such as rock art and musical artifacts, which point to the role of music in ritual, labor, and social bonding.

Music performance serves as a robust context for investigating the integration of cognitive, emotional, and motor functions. The act of performing requires the dynamic coordination of mental and bodily systems, both within individuals and across group settings. Synchronization and adaptation are integral at all stages of performance -- preparation, execution, and post-performance interaction -- highlighting music's function as a deeply embedded, adaptive activity.

Section \ref{cha3:smp:learn} gives an example for adaptation and synchronization in music education (before performance). The preparatory phase of music performance involves the training and refinement of complex sensorimotor and cognitive skills. During this stage, musicians develop adaptive capacities to align motor output with temporal structures, while internalizing stylistic and expressive nuances. Synchronization in this context involves rhythmic entrainment and attentional calibration, facilitating precise timing and anticipation \cite{JUS05, SCH01h}. Through repetitive practice, musicians consolidate these adaptations, establishing the foundation for automatic and flexible performance behavior.

During a performance (Sec.\,\ref{cha3:smp:play}), the execution phase of performance is marked by real-time adaptation to both internal states and external cues. Musicians must continuously monitor auditory and visual information to adjust their output accordingly. Emotional regulation and expressive intent are enhanced through synchronization of autonomic, experiential, and behavioral subsystems \cite{MAU05, GRE07}. In ensemble settings, synchronization extends to inter-performer coordination and conductor-performer dynamics, necessitating mutual adaptation and the negotiation of temporal alignment. This phase exemplifies a complex form of self-organization, where individual contributions are shaped by shared expressive goals.

Post-performance in Sec.\,\ref{cha3:smp:applause}, synchronization mechanisms persist in the interaction between performers and audience members. Applause, as a collective behavioral response, exemplifies interpersonal synchronization driven by shared emotional resonance and social cohesion \cite{EKM92, LEV03}. Research on audience dynamics reveals that clapping patterns often self-organize into temporally coordinated rhythms, reflecting group-level adaptation and emergent social behavior. This underscores the idea that music performance transcends the individual performer, fostering bidirectional communication between performers and listeners.

Music performance, viewed through the lens of adaptation and synchronization, offers profound insights into the interdependencies of cognitive, emotional, and social processes. From the initial stages of practice to the real-time demands of performance and the collective responses of audiences, synchronization operates as a unifying mechanism that enables coordinated action, shared affect, and effective communication. By situating musical performance within a broader framework of embodied and socially situated cognition, we gain a deeper understanding of how music functions as a dynamic, adaptive, and interactive phenomenon.

We now provide a brief overview of the following content: In Sec.\,\ref{sec:model}, we introduce the framework of complex system theory and two foundational mechanisms. Sections~\ref{cha3:smp:learn}, \ref{cha3:smp:play} and \ref{cha3:smp:applause} give a review about research findings in the context of music performance, viewed through the lens of adaptation and synchronization, offers profound insights into the interdependencies of cognitive, emotional, and social processes. Finally, we summarize our conclusions in Sec.\,\ref{sec:conclusion}.

\section{Basic mechanism in the framework of complex system theory}
\label{sec:model}

This section provides an in-depth examination of two foundational mechanisms -- adaptation and synchronization -- that play crucial roles in the dynamics of complex systems. Complex systems, by definition, consist of numerous interdependent components whose interactions lead to the emergence of collective behaviors, such as self-organization and spontaneous order. These behaviors cannot be solely attributed to the individual properties of the components or their nonlinear interactions. Rather, they arise from a dynamic interplay between the individual elements' properties and the network's topology, which together facilitate the emergence of complex patterns \cite{SAW20}. Adaptation and synchronization are central to understanding this emergent behavior and are intertwined in ways that significantly impact the functionality and stability of these systems.

Adaptation within complex systems refers to the capacity of components to modify their behaviors based on the states of other components. This ability to adjust responses in reaction to external stimuli or internal changes is a key factor that enables systems to remain flexible, resilient, and capable of evolving over time. It is through adaptation that complex systems exhibit the capability to reorganize themselves, respond to perturbations, and develop self-organized patterns that support system stability \cite{SAW23b}. Synchronization, on the other hand, refers to the process through which components of a system align their behaviors or states, fostering coordinated activity. Synchronization ensures that diverse elements of a complex system work together cohesively, which is essential for maintaining system-wide coherence and functionality. In many natural and engineered systems, synchronization is a critical mechanism that ensures the system operates efficiently and harmoniously.

\begin{figure}
    \centering
    \includegraphics[width=.6\textwidth]{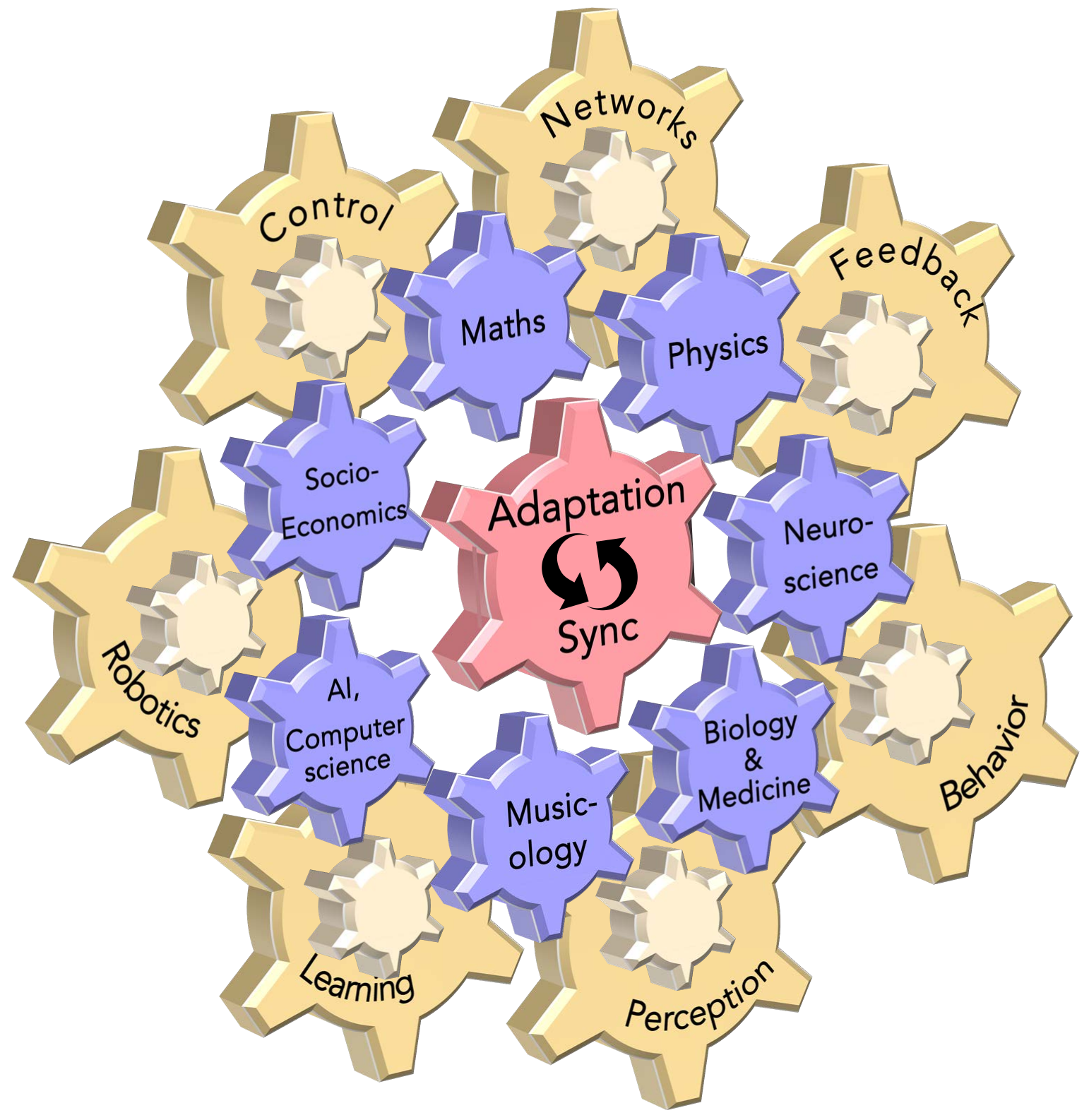}
    \caption{Adaptation and synchronization (``Sync'') across different scientific disciplines (blue) and applications (yellow) as well as their strong interlinking and interlocking, similar to a system of gears. The central rotating gear (red), which symbolizes the drive, shows the interrelationship of the two basic mechanisms, adaptation and synchronization: adaptive mechanisms can lead to synchronization, synchronization in turn can be the motor for adaptive processes.}
    \label{fig:SAW23:gears}
\end{figure}

As shown in Fig.\,\ref{fig:SAW23:gears}, the relationship between adaptation and synchronization is deeply intertwined, with each mechanism influencing and reinforcing the other. Adaptation allows systems to fine-tune their responses in ways that support synchronization, while synchronization facilitates a shared temporal framework that enables further adaptation. Together, these mechanisms form the backbone of complex system dynamics, contributing to phenomena such as the collective behavior observed in biological networks, social systems, and engineered networks. The interplay between adaptation and synchronization is particularly important when considering how complex systems evolve and exhibit behaviors that cannot be reduced to the properties of individual components alone \cite{PIK01,BOC06b}.

Despite significant advances in the study of complex systems, a comprehensive mathematical framework that fully integrates adaptation and synchronization remains elusive \cite{SAW23b}. Current research employs a variety of mathematical approaches to model these mechanisms. For example, differential equations are commonly used to describe the dynamic processes that govern adaptation and synchronization \cite{YAK68,YAK68a}. Graph theory, on the other hand, provides insights into the structural properties of networks, helping to explain how the connectivity between components influences the emergence of collective behaviors \cite{BOC18}. Additionally, stochastic models are often used to account for the uncertainties inherent in complex systems and to explore the probabilistic nature of interactions and outcomes. However, no single model has yet been able to fully encapsulate the breadth of dynamics exhibited by complex systems, highlighting the need for ongoing interdisciplinary research that bridges gaps across fields and integrates diverse approaches.

Synchronization has become a focal point for research due to its ubiquitous presence in both natural and engineered systems. It is a phenomenon that has been observed across a wide range of contexts, from the synchronization of organ pipes \cite{ABE06,ABE09,FIS14,FIS16,SAW18a,SAW20,SAW25} or cardiac pacemakers \cite{GER20} and neuronal firing patterns \cite{SIN18} to the coordinated behavior observed in social systems, such as audience applause \cite{NED00}. In these contexts, synchronization facilitates collective behavior that is critical for the stability and efficiency of the system. Researchers have increasingly focused on understanding how synchronization occurs within networks of oscillators, where factors such as coupling strength, network topology, and the local dynamics of individual components play crucial roles in determining the system's ability to synchronize. These studies have broadened our understanding of how synchronized patterns arise in diverse systems, from simple mechanical systems to the complex neuronal networks of the brain \cite{BOC06b,BOC18}.

One of the most intriguing developments in the study of synchronization is the discovery of partial synchronization patterns, such as chimera states, where synchronized and desynchronized behaviors coexist in the same system. These patterns provide valuable insights into the complexity of synchronization dynamics and have been observed in various types of networks, ranging from ecological systems to social networks \cite{PAN15, SCH16b}. Such partial synchronization reveals the potential for systems to exhibit heterogeneous behaviors while still maintaining overall coherence, challenging traditional models that emphasize complete synchronization. The study of these states has become a key area of research, as they offer new perspectives on how complex systems can balance order and disorder within their dynamics.

In neuroscience, synchronization plays a crucial role in cognitive functions such as attention, perception, and memory consolidation \cite{SIN18}. However, abnormal synchronization is also associated with various pathological conditions, including neurological disorders like epilepsy, where excessive synchronization disrupts normal brain activity \cite{GER20}. This dual role of synchronization -- both as a facilitator of normal brain function and as a contributor to disease -- underscores the importance of understanding the mechanisms governing synchronization in neural networks. The study of synchronization in the brain provides insights not only into fundamental cognitive processes but also into the development of interventions for neurological disorders.

Given the importance of synchronization across various domains, it remains a central topic of research in complex systems science. This review will further explore the roles of adaptation and synchronization in complex systems, offering a detailed analysis of how these mechanisms interact within multilayered systems. By examining concepts such as cluster synchronization, pacemaker-induced chimera states, and relay synchronization, we aim to expand our understanding of how complex systems coordinate behavior across multiple layers and under varying conditions. These studies not only advance the theoretical foundations of complex systems but also provide practical insights for controlling and optimizing networked systems in diverse scientific and engineering applications.


\section{Practicing and flow}
\label{cha3:smp:learn}

The concept of \textit{flow}, introduced by Mih\'aly Cs\'{\i}kszentmih\'alyi in the 1970s, describes a psychological state marked by complete absorption in a task, intrinsic motivation, and optimal performance. Individuals experiencing flow report heightened concentration, loss of self-consciousness, and an altered perception of time. This state arises when the challenge of a task is well-matched to the individual's skill level, facilitating a seamless integration of action and awareness. Cs\'{\i}kszentmih\'alyi first documented this phenomenon among creative professionals, such as painters, who became so engrossed in their work that they ignored basic physiological needs, yet showed reduced interest in the activity once it was completed \cite{NAK21b}.

\subsection{Flow across domains and conceptual challenges}

Flow theory has since found applicability in diverse fields, including education, sports, human-computer interaction, and the performing arts. In music psychology, flow has been instrumental in understanding immersive states during musical practice and performance \cite{MAC06}. Despite the theoretical richness of the concept, empirical research on flow is fragmented. This fragmentation stems from inconsistencies in definitions, measurements, and methodological frameworks, which hinder the development of a coherent and unified model of flow \cite{HAR17b}.

\subsection{Neurocognitive foundations of flow}

Flow states are the product of dynamic interplay between cognitive, neural, and environmental factors. Central to this phenomenon is the balance between task difficulty and personal competence, which facilitates a transition from conscious, effortful processing to automatic and efficient execution \cite{GOL20a}. Two major theoretical models have been proposed to explain the neurocognitive underpinnings of flow:

\begin{itemize}
    \item \textbf{Transient Hypofrontality Hypothesis (THH):} Proposes that flow involves a temporary downregulation of the prefrontal cortex, particularly the dorsolateral prefrontal cortex. This reduction in activity is thought to attenuate self-referential processing and promote task-focused attention \cite{DIE06}.
    
    \item \textbf{Synchronization Theory of Flow (STF):} Suggests that flow emerges from synchronized neural activity between frontoparietal attentional networks and dopaminergic reward systems. This synchronization enhances neural efficiency and facilitates immersion in the task \cite{WEB20}.
\end{itemize}

The basal ganglia play a central role in this process, particularly in the automation of motor and cognitive routines. Dopamine release within these circuits reinforces goal-directed behavior, contributing to the intrinsic rewarding experience of flow. Neural synchronization across the basal ganglia, prefrontal cortex, and parietal areas appears to streamline cognitive operations, reduce perceived effort, and maintain sustained attention.

Emerging evidence points to the potential of non-invasive brain stimulation techniques, such as transcranial direct current stimulation (tDCS), in facilitating flow states. By modulating cortical excitability, particularly in the dorsolateral prefrontal cortex, tDCS has been shown to enhance learning, attention, and skill acquisition in tasks demanding sustained engagement and rapid adaptability. These findings suggest possible applications for optimizing performance in education, rehabilitation, and high-stakes occupational settings.

\subsection{Musical improvisation and flow}

A compelling context for studying flow is musical improvisation, where real-time creativity and motor control intersect with emotional and sensory processing. Neuroimaging research reveals that expert improvisers exhibit a unique neural signature during performance: increased activity in the medial prefrontal cortex, coupled with reduced external monitoring, reflecting a shift toward internally driven, spontaneous expression \cite{ERK18}. This process shares similarities with language production, where structured elements are recombined flexibly and fluently. Improvisation thus provides an ideal model for understanding the dynamic reconfiguration of cognitive systems underlying creative flow states.

Flow continues to gain attention as a psychological and neurophysiological construct with wide-ranging implications. It serves not only as a vehicle for optimizing performance and creativity but also as a coping mechanism for stress and anxiety. However, variability in theoretical and empirical approaches remains a significant challenge. Standardizing methodologies and incorporating neuroscientific tools such as electroencephalography (EEG), functional magnetic resonance imaging (fMRI), and transcranial direct current stimulation (tDCS) may advance our understanding of flow as a synchronized, adaptive, and intrinsically rewarding brain state. The Synchronization Theory of Flow offers a promising integrative framework, positioning flow as a manifestation of optimal neural coupling across attentional and motivational networks, with implications for both basic research and applied settings.

\section{Synchronization and the adaptive function of emotions}
\label{cha3:smp:play}

Emotions have long been conceptualized as complex, patterned, and adaptive physiological responses that facilitate an organism's interaction with environmental and internal demands \cite{LAZ91}. From this adaptive perspective, emotions are not merely subjective states but serve a fundamental role in survival and well-being by preparing the body and mind for appropriate action. This evolutionary-functional view dates back to the foundational work of Darwin \cite{DAR72}, who proposed that emotional expressions evolved to support communication and social coordination.

Subsequent theoretical developments have emphasized that emotions arise through coordinated changes across multiple physiological and neural subsystems, including autonomic responses (e.g., heart rate, respiration), subjective experiences (e.g., feelings), and behavioral expressions (e.g., facial or vocal cues) \cite{SCH01g,MAU05}. A defining feature of emotion is thus the synchronization among these components, a process that enables coherent and functionally adaptive responses \cite{SCH01h}. 

With advances in neuroimaging methodologies, such as positron emission tomography (PET) and functional magnetic resonance imaging (fMRI), researchers have identified distinct neural correlates associated with various emotional states, further underscoring the multidimensional nature of emotional processing \cite{BLO01,KOE14}. These neurophysiological studies demonstrate that emotions are instantiated in distributed brain networks involving limbic, cortical, and brainstem regions, reflecting the integration of cognitive appraisal, bodily feedback, and affective labeling.

\subsection{Music as a context for emotional synchronization}

Music represents a uniquely potent affective stimulus capable of evoking deep emotional experiences that often surpass those elicited by other sensory inputs. These experiences emerge through the dynamic interaction of musical structure, listener disposition, and situational context. Given music's evolutionary roots in social interaction and communication, it provides an ideal domain for studying the synchronization and adaptation of emotional components both within individuals and between social agents.

The synchronization of physiological and neural subsystems in response to music is increasingly recognized as a foundational mechanism for emotional engagement. Music-induced emotions often involve temporal alignment between the body and the rhythmic or harmonic structure of the piece, which facilitates entrainment and affective resonance \cite{GRE07}. This entrainment supports not only the emergence of subjective emotional experiences (or ``feelings'') but also regulatory functions, contributing to stress modulation, mood stabilization, and social cohesion \cite{EKM92,JUS05}.

Although the precise mechanisms underlying emotional synchronization remain under investigation, converging evidence supports the view that a certain degree of coherence among emotional subcomponents is essential for the generation and maintenance of emotional states \cite{MAU05,JUS10a}. In this regard, multi-componential models that examine the integration of physiological, experiential, and expressive dimensions offer a more comprehensive understanding of emotional dynamics, particularly in response to music.

\subsection{Component synchronization and emotional complexity in music}

In the domain of music, synchronization plays a pivotal role not only in shaping emotional experiences but also in enhancing the clarity, expressiveness, and communicative efficacy of musical performance. Emotional reactions to specific musical features -- such as tempo, melody, harmonic tension, and rhythm -- are often mediated by the degree to which bodily and neural systems synchronize with these features. For example, heightened states of surprise, anxiety, or excitement tend to engage stronger and more widespread synchronization patterns, which can be observed at both physiological and cortical levels.

These patterns are often emotion-specific. High-intensity emotions such as awe or fear involve increased coherence among multiple subcomponents and are frequently characterized by rapid feedback mechanisms, resembling the behavior of complex adaptive systems. Importantly, this coherence does not imply uniformity but rather a context-sensitive integration of various processes that enhance the organism's ability to interpret and respond to musical stimuli.

To capture these dynamic interactions, multivariate analytic approaches that assess concurrent physiological signals (e.g., heart rate variability, skin conductance), self-reported affect, and neural activity are increasingly employed. These methods enable the identification of emotion-specific psychophysiological signatures, thereby advancing our understanding of how synchronization mediates the depth and richness of music-induced emotions.

\subsection{Synchronization in musical ensembles}

The generation of emotionally resonant musical performances requires a high level of coordination among ensemble members. Synchronization in this context refers not only to the temporal alignment of musical entries but also to the dynamic interplay of interpretive intentions, bodily movements, and sensory feedback. Successful ensemble performance depends on the precise alignment of individual temporal structures to create a cohesive auditory output.

Interestingly, moderate levels of asynchrony have been shown to enhance the perceptual distinction of musical lines within polyphonic textures, allowing listeners to appreciate individual contributions without perceiving disunity \cite{RAS01a}. This phenomenon suggests that synchronization in musical ensembles is not a binary phenomenon but rather a finely tuned process that balances cohesion with expressiveness.

As ensemble size increases, the challenge of maintaining synchronization grows accordingly. In large ensembles, conductors play a central role in coordinating performance, ensuring temporal unity, and shaping musical interpretation. However, recent research challenges the notion that synchronization is solely the conductor's responsibility. Studies indicate that synchronization is a shared and interactive process, involving auditory and visual cues, internal predictive mechanisms, and mutual adaptation among ensemble members \cite{AUS12,MCP22}.

The conductor's gestures serve as a high-level temporal and expressive framework, but ensemble musicians continuously negotiate timing and expression through both external signals and internal models. This view is supported by studies showing that musicians integrate multiple sources of information -- such as auditory feedback, motor predictions, and co-performer behavior -- to achieve real-time coordination \cite{ONO15}. Thus, synchronization in ensemble performance emerges from a distributed system of mutual attunement rather than a unidirectional top-down control structure.

In sum, synchronization is a critical mechanism underlying both emotional experience and musical performance. Whether manifesting as the integration of emotional subsystems or as the coordinated interplay within musical ensembles, synchronization supports adaptive functioning, social connection, and aesthetic expression. By examining the complex dynamics of synchronization across physiological, neural, and behavioral domains, especially in music, researchers gain valuable insights into the fundamental architecture of emotion and communication.

\section{Interpersonal synchrony}
\label{cha3:smp:applause}

Interpersonal synchrony has emerged as a foundational construct in the study of social interaction, encompassing the alignment of physiological, behavioral, and cognitive processes between individuals exposed to shared stimuli. A substantial body of research emphasizes the central role of synchrony in enhancing mutual understanding, fostering social bonds, and facilitating cooperative behavior. Particularly within the context of musical engagement, recent studies have demonstrated that both motor and physiological synchrony spontaneously occur among audience members during live performances. These phenomena have been correlated with aesthetic appreciation, emotional resonance, and stable personality traits, suggesting that synchronization may serve as a biological substrate for shared affective experiences and interpersonal attunement.

\subsection{Neural and physiological synchronization in social contexts}

A growing number of studies have focused on \textit{inter-brain synchronization} -- the alignment of neural oscillatory activity across individuals during social interactions -- as a key mechanism underpinning human connection. Employing neuroimaging modalities such as electroencephalography (EEG) and related electrophysiological techniques, researchers have documented inter-brain coupling during various interactive behaviors, including collaborative musical performance \cite{MUE21a}, choir singing \cite{DEL23}, joking \cite{TSC23a}, romantic interactions such as kissing \cite{HUG07}, and strategic gameplay like chess \cite{LIU17a}. These studies reveal that synchronous brain activity is not limited to high-affiliation or emotionally charged scenarios but also emerges in structured, cognitively demanding contexts. Such findings support the hypothesis that neural synchronization facilitates the dynamic integration of information, contributing to the emergence of shared mental models, coordinated behavior, and the regulation of joint attention. From a systems neuroscience perspective, these results reinforce the significance of complex systems approaches in investigating distributed neural networks that synchronize during social interaction.

Among the most compelling illustrations of interpersonal synchrony are those observed during live classical music concerts. Studies by Tsakiris and colleagues \cite{TSC23,TSC24} have shown that audience members display significant physiological synchrony -- specifically, alignment in heart rate (HR), heart rate variability (HRV), skin conductance levels (SCL), and respiratory patterns -- when exposed to shared musical experiences. The observed coherence in physiological signals points to an embodied form of cognition, wherein the mind and body dynamically respond to environmental stimuli, particularly music. This embodiment is consistent with the 4E cognition model -- embodied, embedded, enacted, and extended cognition -- which posits that cognitive processes are not confined to the brain but are distributed across the body and the environment \cite{NEW18}. 

Within this framework, physiological synchrony can be understood as a manifestation of continuous feedback loops between internal bodily states and external sensory input, supporting the notion that music operates as a socially situated and embodied phenomenon. Higher levels of synchrony have been associated with reductions in negative affect, increases in positive emotional states, and dispositional traits such as openness to experience. Synchrony varied depending on musical repertoire, with unfamiliar contemporary compositions eliciting the highest levels of physiological alignment \cite{TSC24}. These findings imply that cognitive novelty and attentional engagement play crucial roles in driving interpersonal physiological resonance.

\subsection{Synchronization in musical improvisation}

Synchronization also plays a central role in \textit{musical improvisation}, a highly interactive and temporally constrained form of musical communication. Improvisation requires musicians to spontaneously coordinate their actions in real time, drawing on memory, sensorimotor skills, and creative capacities. The successful execution of improvisation relies on the establishment and maintenance of behavioral, neural, and social synchrony, which enables performers to co-create coherent musical narratives. Research integrating cognitive neuroscience, dynamical systems theory, and audience studies has revealed that synchronization operates across multiple temporal and organizational levels during improvisation, serving as a mechanism for fostering innovation, cohesion, and mutual responsiveness.

Empirical studies using cross-wavelet spectral analysis have uncovered the multiscale dynamics of movement coordination among improvising musicians, demonstrating how performers adapt to each other's timing and gestures over extended periods \cite{ASH15a}. This coordination supports the emergence of \textit{self-organized systems} in which mutual constraints and adaptive feedback loops give rise to novel, emergent musical forms.

Mobile brain-body imaging (MoBI) and electroencephalography (EEG) hyperscanning techniques have shown increased inter-brain synchrony in beta and gamma bands ($13$--$50$ Hz) during live jazz performances \cite{RAM23}. These oscillations are linked to anticipatory processing, shared intentions, and real-time adaptation \cite{NEL22}. Synchrony is not limited to performers; audiences also display spontaneous motor entrainment with performers and each other \cite{TAK23}, contributing to a shared immersive experience described as group flow.

Live accompaniment further enhances synchronization, reward sensitivity, and engagement during improvisation \cite{PAL24}. It supports social cooperation and mutual responsiveness, amplifying the collaborative aspects of music-making.

\subsection{Collective synchronization in audience behavior}

Beyond performers, synchronization is also critical in shaping collective audience behavior. Research by Nomura et al. \cite{NOM15,NOM24} on heart rate synchronization during music listening has revealed that intra-individual synchrony exceeds inter-individual alignment, indicating that internal consistency in physiological responses is a more robust predictor than external emotional states or musical preferences.

Applause provides another compelling example of self-organized synchronization. Studies have shown that applause begins as uncoordinated clapping, which transitions into a synchronized rhythmic phase through \textit{period doubling} \cite{NED00}. This shift reflects principles of globally coupled oscillators, where reduced frequency dispersion facilitates alignment.

These dynamics have been empirically confirmed in both field settings and experimental contexts. However, synchronization is often disrupted when individuals attempt to increase volume by shortening clap intervals, revealing a tension between collective order and individual expression.

Further studies on applause in academic settings confirm that both the initiation and cessation of clapping are governed by social contagion effects \cite{MAN13b}. Using Bayesian models, it was shown that individuals are more likely to begin or stop clapping based on the behavior of others, rather than spatial cues. This reveals that applause is less about individual appreciation and more about socially regulated behavior patterns.

Taken together, this growing body of research highlights the multifaceted role of synchronization in shaping human interaction. From neural oscillations and physiological coupling to collective behaviors such as clapping and musical collaboration, synchronization emerges as a foundational principle underlying complex social dynamics. Its presence in both passive and active contexts underscores its adaptability and centrality to human cognition, communication, and culture.

\section{Conclusion} 
\label{sec:conclusion}

The field of music performance represents a dynamical interplay among musicians, the music itself, and listeners. Moreover, music performance serves as a rich domain for examining adaptation and synchronization, revealing their central roles in shaping the cognitive, emotional, and social facets of musical expression. By analyzing the interconnected stages of preparation, execution, and reception, interdisciplinary research in the interface of music psychology and complex systems offers profound insights into the mechanisms driving musical creativity and collective engagement.

This review has made use of the key concepts of adaptation and synchronization from complex systems theory and explores their relevance to music performance research. Complex systems, characterized by the interactions of numerous interdependent components, give rise to emergent behaviors such as self-organization and coordination. Adaptation allows components to modify their responses based on others, while synchronization facilitates coordinated behavior across a network. These mechanisms are fundamental for understanding collective behaviors in both natural and engineered systems. This review aimed to transfer these concepts to the study of music performance, where adaptation and synchronization are crucial for understanding the dynamic interactions between performers, their environment, and the audience. By drawing parallels between complex systems and music performance, we seek to enrich our understanding of the coordination, creativity, and collective behaviors that emerge during musical performances.

Flow exemplifies a dynamic state of neurocognitive adaptation, in which synchronization across brain systems -- particularly attention, motor control, and reward circuits -- supports focused engagement and peak performance. The transition from controlled to automatic processing reflects the system's ability to adaptively reorganize, enabling efficient task execution and sustained motivation. Understanding flow as a product of neural synchronization offers valuable insights into the adaptive functions of cognition in high-demand contexts.

Emotional and musical experiences are shaped by synchronization across physiological and neural subsystems. Music-induced emotions rely on coherent interactions among cognitive, affective, and sensorimotor components. In ensemble performance, synchronization emerges through shared responsibility among musicians, mediated by auditory, visual, and predictive cues. Together, these findings highlight synchronization as a fundamental mechanism linking emotion, adaptation, and social interaction.

Interpersonal synchronization -- encompassing physiological, behavioral, and neural alignment -- plays a fundamental role in social and musical contexts. Studies show that shared musical experiences, such as concerts or improvisation, elicit synchronized heart rate, respiration, and brain activity among both performers and audiences. These phenomena are linked to emotional engagement, personality traits, and cognitive processes, supporting theories like 4E cognition (embodied, embedded, enactive and extended cognition). In improvisation, synchronization facilitates creativity, coordination, and group cohesion, while audience behaviors such as applause exhibit self-organized rhythmic patterns governed by social contagion and complex systems dynamics. Overall, synchronization emerges as a core mechanism underpinning collective experience and social bonding.

Concluding, music performance constitutes a dynamic interaction between musicians, music, and listeners, offering a unique framework for investigating core principles of complex systems. This review highlights the relevance of adaptation and synchronization -- two foundational mechanisms in complex systems theory -- to music performance research. By examining the interconnected phases of preparation, execution, and reception, the study explores how these mechanisms shape the cognitive, emotional, and social dynamics of musical expression. Adaptation facilitates flexible responses to internal and external cues, while synchronization enables coordinated behavior within and between individuals. Insights from flow states, emotional processing, and interpersonal synchrony further underscore the role of synchronized neural and physiological processes in fostering musical creativity, engagement, and social cohesion. This interdisciplinary approach bridges complex systems theory and music psychology, offering a deeper understanding of the mechanisms underlying musical interaction and collective experience.

\section*{Acknowledgments}
The author acknowledge Matthias Bertsch, Isabel Fernholz, and Gary McPherson for their valuable contributions through insightful discussions. Special appreciation is extended to the Berlin University of the Arts for providing an inspiring environment in which many of the concepts presented in this work were further developed, particularly during the ``creative performance workshop'' organized by Isabel Fernholz and Jakub Sawicki in 2024.

\bibliography{ref}
\bibliographystyle{tfs}

\end{document}